\documentclass[12pt, a4paper]{article}
\usepackage{graphicx}

	\newcommand{\ket}[1]%
	{\left|#1\right\rangle}
	\newcommand{\bra}[1]%
	{\left\langle #1\right|}

\title{GRANIT project: a trap for gravitational quantum states of UCN}
\date{}

\begin{document}
\maketitle

\author{
\begin{center}

V.~V.~Nesvizhevsky, A.~K.~Petukhov, H.~G.~B{\"o}rner, T.~Soldner, P.~Schmidt-~Wellenburg, M.~Kreuz \\
Institut Laue-Langevin, Grenoble, France \\
\bigskip
\and

\underline{G.~Pignol}, K.~V.~Protasov, D.~Rebreyend, F.~Vezzu \\
LPSC, UJF/CNRS-IN2P3/INPG, Grenoble, France \\
\bigskip
\and

D.~Forest, P.~Ganau, J.~M.~Mackowski, C.~Michel, J.~L.~Montorio, N.~Morgado, L.~Pinard, A.~Remillieux \\
Laboratoire des Mat\'eriaux Avanc\'es, Villeurbanne, France \\
\bigskip
\and

A.~M.~Gagarski, G.~A.~Petrov \\
Petersburg Nuclear Physics Institute, Gatchina, Russia \\
\bigskip
\and

A.~M.~Kusmina \\
Khlopin Institute, St.Petersburg, Russia \\
\bigskip
\and

A.~V.~Strelkov \\
Joint Institute of Nuclear Research, Dubna, Russia \\
\bigskip
\and

H.~Abele \\
University of Heidelberg, Germany \\
\bigskip
\and

S.~Bae{\ss}ler \\
University of Mainz, Germany \\
\bigskip
\and

A.~Yu.~Voronin \\
Lebedev Physical Institute, Moscow, Russia
\end{center}
}

\newpage

\begin{abstract}
Previous studies of gravitationally bound states of ultracold neutrons
showed the quantization of energy levels, and confirmed quantum
mechanical predictions for the average size of the two lowest
energy states wave functions. Improvements in position-like
measurements can increase the accuracy by an order of magnitude only.
We therefore develop another approach, consisting in
accurate measurements of the energy levels. The GRANIT experiment
is devoted to the study of resonant transitions between quantum states
induced by an oscillating perturbation.

According to Heisenberg's uncertainty relations, the accuracy of
measurement of the energy levels is limited by the time available
to perform the transitions. Thus, trapping quantum states will be necessary,
and each source of losses has to be controlled in order to
maximize the lifetime of the states. We discuss the general
principles of transitions between quantum states, and consider
the main systematical losses of neutrons in a trap.
\end{abstract}

%--------------------%
\section{Introduction}
%--------------------%

As predicted by quantum mechanics, a neutron bouncing above a mirror has discrete energy states.
Due to the extreme weakness of gravity, these quantum states exhibit outstanding properties.

First, the mean height of the $n^{\rm th}$ state is much larger than atomic size,
\begin{equation}
z_n \approx (n-1/4)^{2/3} \times 11 \ \mu{\rm m},
\end{equation}
which has been used to prove the quantization of energy at the ILL high flux reactor in Grenoble \cite{Nesvizhevsky:2002, Nesvizhevsky:2003}.
A neutron flux has been measured through a narrow slit between a horizontal mirror and a scatterer above.
We observed the discrete behaviour of this flux as a function of the height of the slit.
In particular, the flux is zero if the height is less than $10 \ \mu {\rm m}$, since no quantum state can penetrate throught the slit. 
The classical turning points of the two first states have then been determined to be \cite{Nesvizhevsky:2005ss},
\begin{eqnarray}
z_1^{{\rm exp}} & = & 12.2 \pm 1.8_{\rm sys} \pm 0.7_{\rm stat} \ \mu {\rm m}, \\
\nonumber
z_1^{{\rm th}} & = & \frac{3}{2} \bra{1}\hat{z}\ket{1} \ = \ 13.7  \ \mu {\rm m},
\end{eqnarray}
\begin{eqnarray}
z_2^{{\rm exp}} & = & 21.6 \pm 2.2_{\rm sys} \pm 0.7_{\rm stat} \ \mu {\rm m}, \\
\nonumber
z_2^{{\rm th}} & = & \frac{3}{2} \bra{2}\hat{z}\ket{2} \ = \ 24.0  \ \mu {\rm m}.
\end{eqnarray}
Higher states are more difficult to resolve.
We expect to be able to improve the precision of position-like observables by one order of magnitude at most.

One second important property of the quantum states is the extreme smallness of the energies. 
Indeed, the energy of the $n^{\rm th}$ state is, within the Bohr-Sommerfeld approximation, 
\begin{equation}
E_n \approx (n-1/4)^{2/3} \times 1.7 \ {\rm peV}.
\end{equation}
Then, the corresponding frequencies $f_n = \frac{E_n}{h}$ are small as well.
They are in the kilohertz range, which makes it possible to probe the quantum states with an oscillating perturbation at these easily accessible frequencies. 
In the next section, we will discuss the principle of the GRANIT experiment, which aims to induce resonant transitions between quantum states.
The accuracy of the transition frequencies is limited by the lifetime in the quantum levels.
To increase the duration of the perturbation, a trap has to be built, 
and the third section is devoted to the estimation of lifetimes of quantum states in this trap.

%-----------------------------------------------------%
\section{Resonant transitions in the GRANIT experiment}
%-----------------------------------------------------%

\begin{figure}
\begin{center}
\includegraphics[width=.8\linewidth]{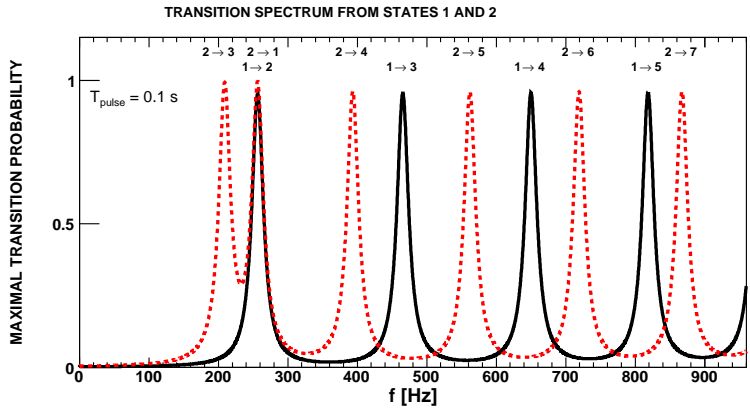}
\caption{Maximum transition probability 
for neutrons prepared in the first quantum state (straight line) 
and for neutrons prepared in the second quantum state (dotted line), as a function of the perturbation frequency.
The Rabi angular frequency $\Omega$ satisfies $\frac{\pi}{\Omega} = 0.1 \ {\rm s}$.}
\label{transition}
\end{center}
\end{figure}

Let us assume that a given neutron bouncing above a horizontal mirror stands in a pure initial state $\ket{N}$ concerning its vertical motion.
We apply a periodic perturbation,
\begin{equation}
\hat{V}(t) = Re \left( V(z) e^{i \omega t} \right),
\end{equation}
induced by an oscillating magnetic gradient, oscillations of the horizontal mirror itself, or by the motion of neighbouring mass.
If the angular frequency $\omega$ is close to $\frac{E_n-E_N}{\hbar}$, which corresponds to the transition to an excited state $\ket{n}$, then a Rabi resonance is expected.
The probability to observe this transition is well known,
\begin{equation}
\label{resonanceformula}
P_{N \rightarrow n}(t) = \frac{1}{1 + \left( \frac{\omega - \omega_{N n}}{\Omega_{N n}} \right)^2 } \sin^2 \left( \sqrt{(\omega - \omega_{N n})^2 + \Omega_{N n}^2} \ \frac{t}{2} \right),
\end{equation}
where  $\Omega_{N n} = \frac{2}{\hbar} \bra{n} V(z) \ket{N}$ is the Rabi frequency, 
defining the perturbation strength for the $N \rightarrow n$ transition.
The maximum of the transition probability,
\begin{equation}
P_{N \rightarrow n}^{\rm max} = \frac{1}{1 + \left( \frac{\omega - \omega_{N n}}{\Omega_{N n}} \right)^2 }
\end{equation}
has a Lorentzian shape, and is reached if the perturbation is applied for the so-called \emph{pulse time},
\begin{equation}
\label{pulsetime}
T_{\rm pulse} = \frac{\pi}{\sqrt{(\omega - \omega_{N n})^2 + \Omega_{N n}^2}}.
\end{equation}
Fig. \ref{transition} shows the maximum transition probability of the first two quantum states for frequencies below $1$ kHz.

Obviously, the pulse time must be smaller than the storage time of neutrons in a given quantum state.
Thus, we can deduce from the previous discussion the two fundamental reasons to increase the storage time.
First, the resonance formula (\ref{resonanceformula}) contains the Heisenberg's uncertainty relation, $\Delta E \cdot T > h/2$, 
where $T$ is the storage time ($T > T_{\rm pulse}$) and $\Delta E$ the width of the resonance curve. 
As a work hypothesis we consider that the accuracy of GRANIT to measure energy transitions equals this resonance width, which is then inversely proportional to the storage time.
In expression (\ref{pulsetime}) we can see the second reason to increase the storage time, 
since the perturbation strength needed for a $100\%$ probability at the resonance is inversely proportional to the pulse time as well.
In the remaining of this section we will explicit this relation connecting strength of perturbation and storage time in the case of magnetically induced transitions, 
and in the next section, we will estimate the lifetime of the quantum states in the GRANIT trap, which is of primary importance.

For now, let us list the characteristic timescales of the problem:
\begin{enumerate}
\item The pulse time needed to resolve the states is about $10$ ms, leading to a resonance curve with a width of half the separation frequency between neighbouring states.
\item The flow through mode time is about $75$ ms, which is the time a neutron of horizontal velocity $4$ m/s takes to pass above a $30$ cm long horizontal mirror.
Of course this time could be increased with slower neutrons or a longer mirror, but we need to trap neutrons to gain orders of magnitude.
Let us notice, however, that it is possible to resolve the states in flow through mode.
\item The ultimate storage time is given by the $\beta$ decay lifetime of the neutron, i.e.\ $886$ s. 
If neutrons can be trapped in a given quantum level that long, energy levels could be measured with a relative accuracy of about $10^{-6}$.
\item We may consider the possibility of radiative decays of the quantum states by spontaneous graviton emissions. 
Nonetheless the corresponding time has been found to exceed the age of the universe by several orders of magnitude \cite{Pignol:2007}, and can be completely neglected.
Therefore, contrary to quantum levels in atomic or nuclear physics, the gravitational quantum levels are absolutely stable, except for external perturbation or the decay of the neutron itself.
\end{enumerate}

%------------------------------------------------%
\subsection{Transitions induced by magnetic field}

Let us examine how strong this perturbation has to be in the case of magnetically induced transitions. 
Since an uniform magnetic field does not couple different quantum states,  we apply the magnetic \emph{gradient}
\footnote{This field actually violates Maxwell's equations. We may add the gradient term $\left( \beta_x {\bf e}_z - \beta_z {\bf e}_x \right) x \cos(\omega t)$, but this term does not couple different quantum states.}
\begin{equation}
{\bf B} = \left( \beta_z {\bf e}_z + \beta_x {\bf e}_x \right)   z \cos(\omega t),
\end{equation}
whose variation with time induces two different transitions.
The first one does not change the spin-state and is induced by the gradient of the vertical component,
\begin{equation}
\hat{V}(t) = - \hat{\mu}_z \beta_z \ \hat{z} \ \cos(\omega t),
\end{equation}
where $\mu$ is the magnetic moment of the neutron.
Contrary, the second transition is induced by the gradient of the horizontal component and changes the spin-state of the neutron,
\begin{equation}
\hat{V}_{\rm flip}(t) = - \hat{\mu}_x \beta_x \ \hat{z} \ \cos(\omega t).
\end{equation}
This property of spin-flipping is important since it can be used to detect that a transition has occured.
In any case, the Rabi angular frequency is given by
\begin{equation}
\Omega_{N \rightarrow n}^{\rm magnetic} = \frac{2}{\hbar} \mu \ \beta \ \bra{n} \hat{z} \ket{N}.
\end{equation}
\begin{figure}
\begin{center}
\includegraphics[width=.7\linewidth]{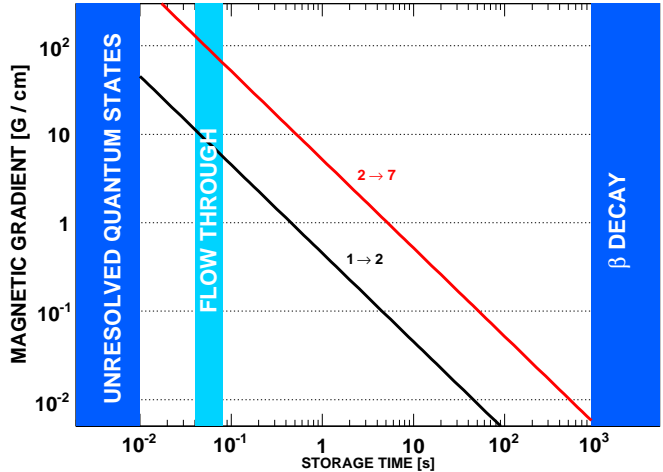}
\caption{Magnetic field gradient needed for a $100\%$ transition probability at resonance as a function of the time during which the transition is performed, 
for the $1 \rightarrow 2$ and the $2 \rightarrow 7$ transition.
Also shown is the minimum time needed to resolve the quantum states, the ultimate $\beta$ decay time, and the typical time in flow through mode.}
\label{gradient}
\end{center}
\end{figure}
Fig. \ref{gradient} shows the magnetic gradient needed for a $100\%$ transition probability at the resonance as a function of the pulse time.
If a storage time of $10$ s is achieved, then we can use a magnetic gradient as low as 0.01 T/m to induce the $2 \rightarrow 7$ transition.
In flow through mode, a 100\% probability for the $1 \rightarrow 2$ transition can
be induced by a magnetic gradient of 0.1 T/m. By using the spin-flip,
the sensitivity to detect the transition can be increased significantly
permitting to observe transitions even with smaller gradients.

%-----------------------------------------%
\subsection{Effects of the Earth' rotation}

Besides the $\beta$ decay lifetime of the neutron, there is another (almost) unavoidable effect limiting the precision of the transitions energies.
Indeed, the rotation of the Earth induces non-inertial effects, described by the potential $- {\bf \Omega}_{\rm Earth} \cdot {\bf \hat{L}}$, 
where ${\bf \Omega}_{\rm Earth}$ is the rotation vector of the Earth, and ${\bf \hat{L}}$ is the angular momentum of the neutron relatively to the Earth' center. 
The main effect induced by this potential is the coupling between horizontal velocity and vertical motion,
\begin{equation}
V_{\rm Earth \ rotation} = - \Omega_{\rm Earth} \cos \theta  \ m v_{NS} \ \hat{z}
\end{equation}
where $v_{NS}$ is the neutron velocity in North-South direction and $\theta$ is the latitude (in Grenoble $\cos \theta = 0.7$).
Thus, the effective vertical acceleration of the neutrons is shifted, depending on its horizontal velocity.
For the first energy state, we can estimate the shift between neutrons travelling in North-South direction and neutrons travelling in East-West direction, for an absolute velocity of $5$ m/s,
\begin{equation}
\left( \frac{\Delta E}{E_1} \right)_{\rm Earth \ rotation} \approx 10^{-6}.
\end{equation}
Figure \ref{lifetimes} shows the corresponding frequency shifts as a function of the state number.
This effect is close to the ultimate sensitivity for energy levels measurements for the ground state, and is even higher for excited states.
Let us notice that Earth rotation also induces a Zeeman shift between spin-up and spin-down states, but this effect is much smaller than the ultimate sensitivity,
\begin{equation}
\left( \Delta E \right)_{\rm Earth \ Zeeman} = \hbar \Omega_{\rm Earth} \approx 6 \times 10^{-8} \ {\rm peV}.
\end{equation}

%----------------------------------------------%
\section{Lifetimes of quantum states in the trap}
%----------------------------------------------%

\begin{figure}
\begin{center}
\includegraphics[width=.8\linewidth]{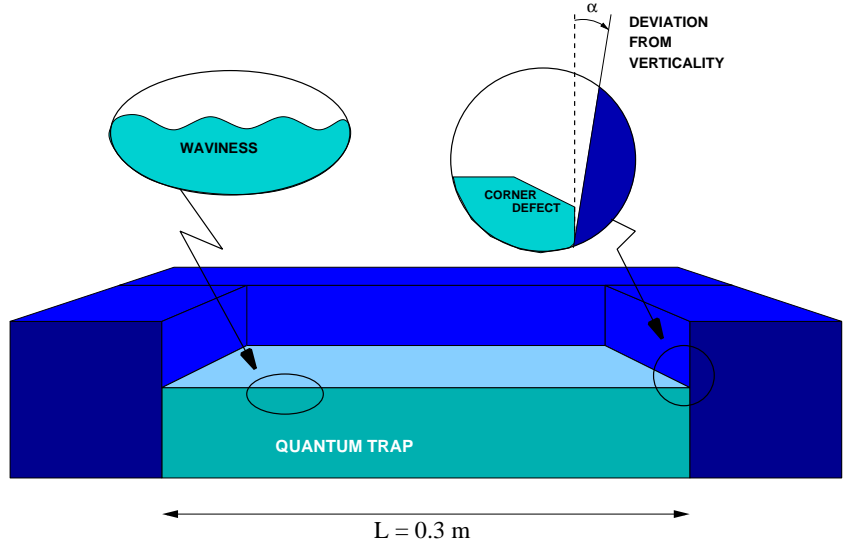}
\caption{Scheme of the GRANIT trap for gravitational quantum levels of UCN.
It shows the main geometrical sources for losses of neutrons: 
the waviness of the bottom mirror, the deviation from verticality of the side wall, and the corner defect.}
\end{center}
\label{trap}
\end{figure}

The trap of quantum states in the GRANIT setup will look like in fig. \ref{trap}.
It is a $30$ cm square bottom horizontal mirror surrounded by vertical side walls.
The horizontal velocity of neutrons in the trap will be about $5$ m/s, this velocity is limited by the Fermi potential of the side walls.
In this section, we estimate the loss rate of neutrons due to geometrical imperfections of the trap as shown in fig. \ref{trap}, that is, the waviness of the bottom mirror, the deviation from verticality of the side walls, and the corner defects.
Other sources of losses, such as 
the seismic noise, 
a possible remaining inhomogeneous static magnetic field, 
interaction with magnetic impurities under the surface, 
interaction with dust or a hydrogen layer on the surface, 
interaction with low-energy phonons in the mirror and 
diffuse scattering, 
will not be considered here.

%------------------------------------------------------%
\subsection{Losses due to waviness of the bottom mirror}

Let us assume that the bottom mirror has a wavy profile $\xi(x)$, so that a neutron with horizontal velocity $v$ sees in its rest frame a time-dependant boundary $\xi(v t)$.
We aim to calculate the loss rate due to the transfer of horizontal velocity to vertical motion.
This problem has already been solved in the case of seismic noise \cite{protasov}, and will be developed in more details in a forthcoming publication. 
Following this perturbation approach, it can be shown that the rate for the $N \rightarrow n$ loss channel reads
\begin{equation}
\Gamma_{N\rightarrow n}^{\rm wavy} = \left( \frac{m g}{\hbar} \right)^2 \frac{1}{v} PSD \left( \frac{f_{N n}}{v} \right)
\label{wavyloss}
\end{equation}
where $f_{N n}$ is the frequency of the $N \rightarrow n$ transition and $PSD(K)$ is the power spectral density of spectral noise,
\begin{equation}
PSD(K) = \lim_{L \rightarrow + \infty} \frac{1}{L} \left| \int_{0}^{L} \xi(x) e^{2 i \pi K x} dx \right|^2.
\end{equation}
The GRANIT bottom mirror has not been nuilt yet, but the power spectral density of a high-quality $300$ mm Si substrate has been measured with several characterisation methods \cite{Assoufid},
\begin{equation}
PSD(K) = \left( \frac{K}{1 \ {\rm mm}^{-1}} \right)^{-2.9} \ 2 \times 10^{-4} \ {\rm nm}^2 \ {\rm mm}.
\end{equation}
It is thus possible to estimate the loss rate of a given quantum level, summing all final states in eq. (\ref{wavyloss}).
The result is shown in fig. \ref{lifetimes}, as a function of initial state quantum number, and for a horizontal velocity of $5$ m/s. 
For the first $30$ quantum states of interest, the rate of losses due to the waviness of the bottom mirror is much smaller than the $\beta$ decay rate.

%--------------------------------------------------------------------%
\subsection{Losses due to deviation from verticality of the side wall}

\begin{figure}
\begin{center}
\includegraphics[width=.8\linewidth]{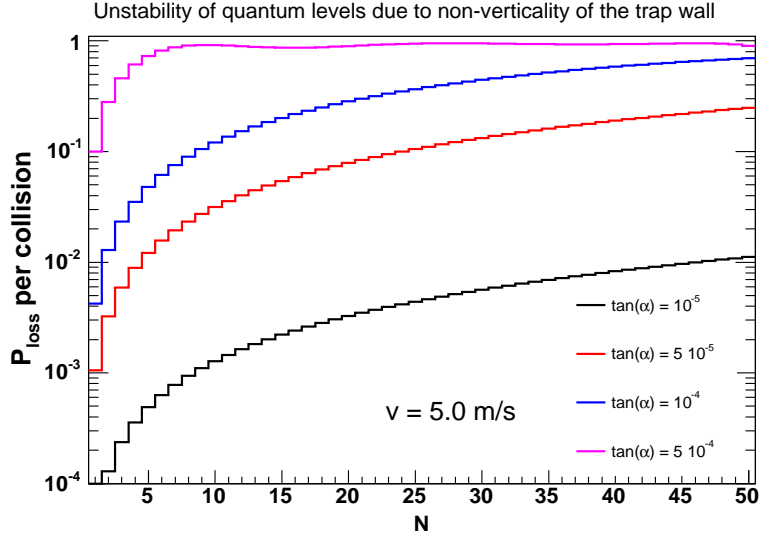}
\caption{Escape probability from quantum states while colliding at the side wall as a function of the quantum state number, 
for different values of the angle $\alpha$ between side wall and vertical axis.} 
\label{verticality}
\end{center}
\end{figure}

A vertical side wall is obviously a necessary feature of the GRANIT trap. 
Here we estimate how precisely vertical this wall has to be.
The probability of transition to different states due to deviation from verticality of the side walls can be calculated in the framework of 
the "sudden kick" approximation. 
Indeed, the collision with a wall can be treated as a sudden kick, during which the neutron gets the vertical momentum $k \sin(2 \alpha)$, 
where $\alpha$ is an angle between the wall and the vertical direction,
 $k = m v / \hbar$ is the wave number of horizontal motion. 
Following the solution given by A.B. Migdal (see e.g. \cite{landau}), we immediately get the probability to leave a given state $\ket{N}$ during a collision,
\begin{equation}
P_N^{\rm wall} = 1 - \left|\bra{N} \exp(i k \sin(2 \alpha)z) \ket{N} \right|^2
\end{equation}
The result is shown in fig. \ref{verticality}, for neutrons with horizontal velocity of $5$ m/s, and for different values of $\alpha$.
Notice that this probability is a quadratic function of $v \ \alpha$ (for small $\alpha$).
As shown in fig. \ref{lifetimes}, the corresponding loss rate $\Gamma_N^{\rm wall} = \frac{v}{L} P_N^{\rm wall}$ (for $\alpha = 10^{-5}$ rad and $v = 5$ m/s) 
is comparable to the $\beta$ decay rate for the very first quantum states, and is more than ten times larger for state number bigger than 10.
We conclude that specific efforts have to be undertaken to set the verticality at an accuracy better than $10^{-5}$.
This will be challenging, because the height of the side mirror is as low as $\approx 1$ mm, but such a precision seems possible.

%---------------------------------------------%
\subsection{Losses due to defects in the edges}

\begin{figure}
\begin{center}
\includegraphics[width=.8\linewidth]{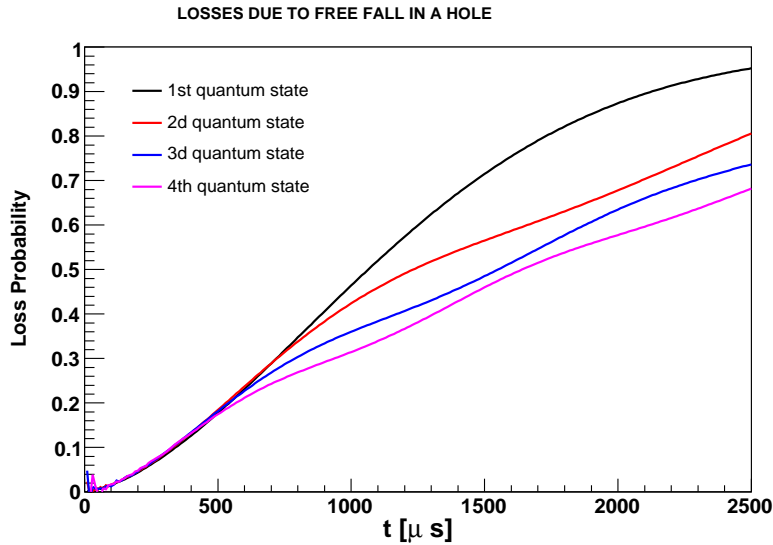}
\caption{Probability for losing quantum states due to free fall as a function of free fall time, for the first quantum states.} \label{hole}
\end{center}
\end{figure}

\begin{figure}
\begin{center}
\includegraphics[width=.8\linewidth]{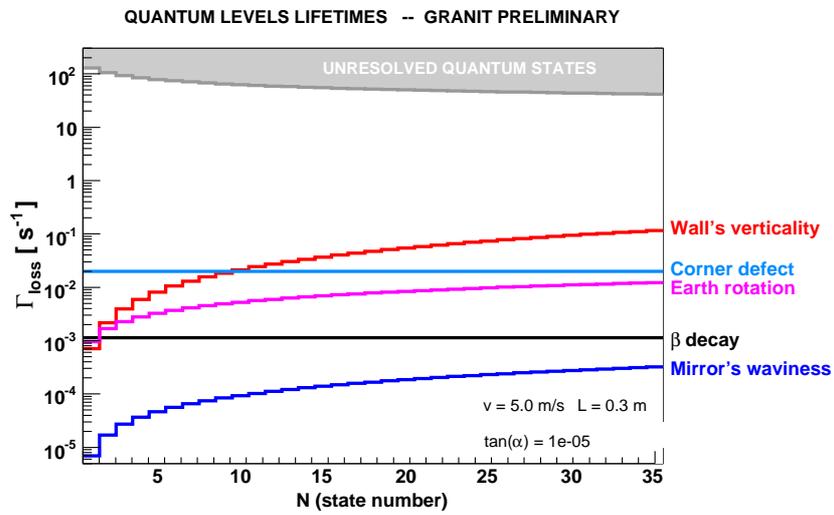}
\caption{Loss rate of trapped quantum states as a function of the level number, due to effects analysed in the text: 
$\beta$ decay, waviness of the bottom mirror, free fall in the corner defect, deviation from verticality of the side walls.
Also shown the minimum loss rate needed to resolve the state, and minimum loss rate needed to be sensitive to Earth rotation bluring of the energy levels.} 
\label{lifetimes}
\end{center}
\end{figure}

The bottom mirror edges cannot be perfectly flat, we will consider the realistic case where a $50 \ \mu$m brink, considered here as a hole, is present for the whole boundary.
Neutrons can either be lost in this hole, or be reflected by the hole in a different quantum state.
To estimate the probability of leaving the initial state $\ket{N}$ during the collision at the corner, we assume a free fall evolution (without bottom mirror) during the classical time:
\begin{equation}
t_{\rm free \ fall} = \frac{2 \times 50 \ \mu{\rm m}}{5 \ \rm{m/s}} = 20 \ \mu{\rm s},
\end{equation}
the effective size of the hole is twice the geometrical size due to reflection on the vertical mirror.
Since the propagator of free fall evolution is well known, the wave function at time $t_{\rm free \ fall}$ reads:
\begin{equation}
\psi(z, t_{\rm free \ fall}) = \left( \frac{m}{2 i \pi \hbar t} \right)^{1/2} \int e^{i \frac{m}{2 \hbar} \left( \frac{(z - z')^2}{t} - (z + z') g t\right)} \psi(z', 0) dz'.
\end{equation}
The amplitude for the neutron to be reflected in the same quantum state $\ket{N}$ 
is thus given by the overlap of this evolved wave function with the wave function corresponding to $\ket{N}$:
\begin{equation}
A_N(t) = \int \ \psi_N(z) \psi(z, t_{\rm free \ fall}) dz
\end{equation}
The loss probability, $1-|A_N|^2$, is shown in fig. \ref{hole} as a function of the falling time, for the first quantum states.
This figure shows clearly that the quantum states are completely lost after an elapsed time corresponding to $10 \ \mu {\rm m}$ free fall $\sqrt{2 \cdot 10 \ \mu{\rm m}/g} = 1 \ {\rm ms}$.
It also shows that for smaller times, all quantum states have the same probability to be lost, that is, $P^{\rm corner} \approx 10^{-3}$ for $t_{\rm free \ fall}$.
As shown in fig. \ref{lifetimes}, the related loss rate $\Gamma_N^{\rm corner} = \frac{v}{L} P^{\rm corner}$ 
is ten times larger than the $\beta$ decay rate.
To reach the ultimate sensitivity, specific efforts have to be undertaken to minimize the corner defects.
However, even in the pessimistic situation presented above, corner defects do not forbid $10$ s storage time for trapped quantum levels.

%------------------%
\section{Conclusion}
%------------------%

The GRANIT experiment will measure the transition energies of gravitationally bound quantum states of neutrons below the kilohertz range, 
by inducing resonant transitions.
The commissionning phase of GRANIT will start in April, 2008, and the first measurements are expected in 2009.
We argued that increasing the storage time of trapped neutrons is an essential feature of the GRANIT experiment, 
since the width of the resonance curve is inversely proportional to the pulse time.
We showed that magnetically induced transitions are doable even in flow through mode, 
and it will be possible to resolve the first resonances in the very first stage of the GRANIT experiment.
The accuracy of the transition energies is limited by the $\beta$ decay lifetime, 
and by the shift in energies due to noninertial effects induced by Earth rotation.
This last subtle effect can in principle be avoided using only neutrons travelling in the East-West direction, 
but this will not be necessary since this effect will be dangerous only if we approach the ultimate sensitivity.
Then the lifetime of trapped quantum levels due to imperfections of the trap were estimated, and compared to the ultimate $\beta$ decay storage time.
We showed that the waviness of the bottom mirror is by no means a problem.
The main source of losses comes from the side walls of the trap: both the deviation from verticality and the corner defects have to be minimized.
Even in the most pessimistic case, a storage time of $10$ s in the trap can be reached, corresponding to the relative accuracy $10^{-4}$ of transition energies.

%-------------------------%
\section*{Acknowledgements}
%-------------------------%

We are grateful to the French Agence Nationale de la Recherche (ANR) for supporting this project.

%------------------------%

\end{document}